# From Prompts to Worlds: How Users Iterate, Explore, and Make Sense of AI-Generated 3D Environments

Aung Pyae*

International School of Engineering, Faculty of Engineering, Chulalongkorn University, Bangkok, Thailand,

aung.p@chula.ac.th

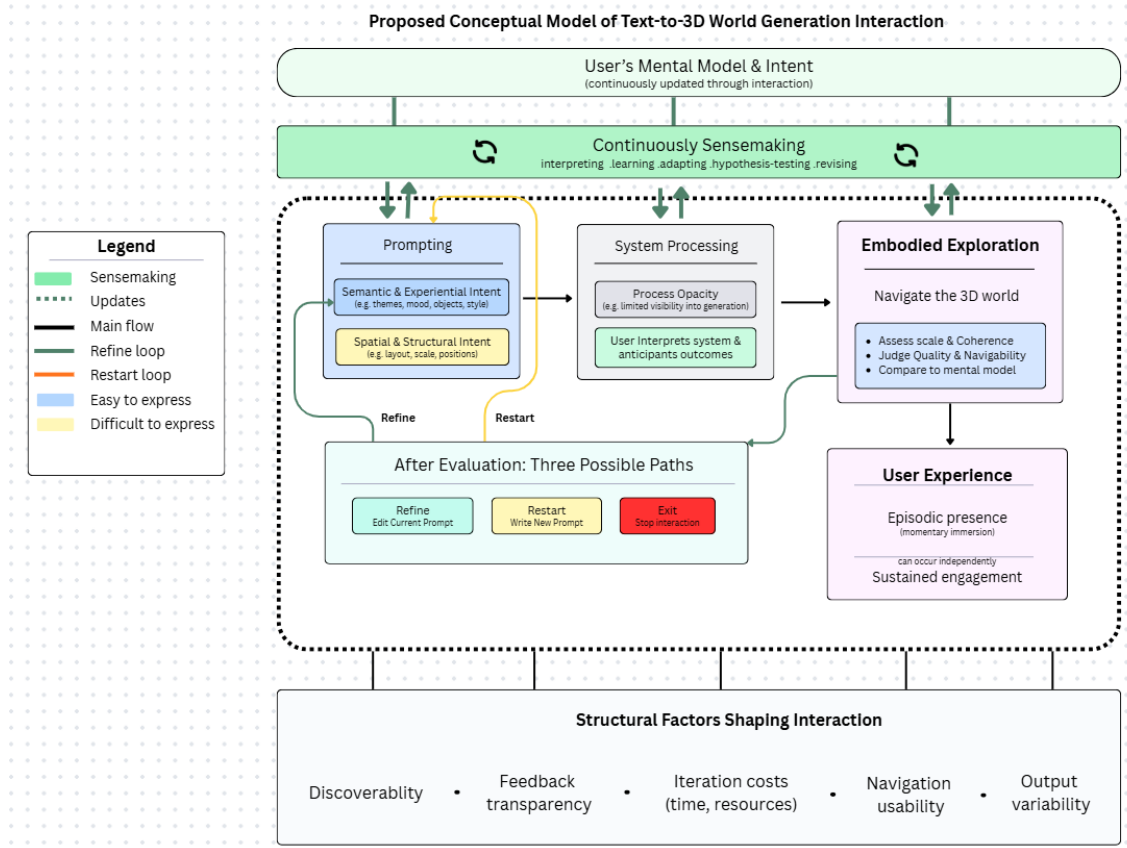

Figure 1: Conceptual model of text-to-3D world generation as iterative sensemaking. Users externalize mental models through prompts expressing semantic intent (easy to articulate) and spatial intent (difficult to specify). The system produces probabilistic 3D worlds under processing opacity, evaluated through embodied exploration. After evaluation, users refine, restart, or exit. Structural factors—

---

* Place the footnote text for the author (if applicable) here.

discoverability, feedback transparency, iteration costs, navigation usability, output variability—shape whether iteration sustains or breaks down. Experience manifests as episodic presence with sustained engagement rather than continuous immersion

Text-to-3D generative AI systems create navigable environments from natural language prompts, but unlike text-to-image generation, evaluation requires embodied exploration of spatial coherence, scale, and navigability. We present the first empirical study of a commercial text-to-3D platform (N = 24), combining think-aloud protocols, behavioral observation, and validated measures of usability, presence, and engagement. We report three findings. First, asymmetric expressibility: users readily convey semantic intent (themes, atmosphere) but struggle to specify spatial structure (layout, scale), reflecting a language-to-space limitation rather than a skill deficit. Second, episodic presence: immersion arises when expectations align with outputs but does not accumulate into sustained place illusion. Third, structural iteration breakdowns: refinement fails due to interaction barriers—poor discoverability, opaque feedback, and high temporal costs—rather than user limitations. Together, these dynamics form a reinforcing cycle in which spatial mismatches persist, producing episodic presence and ongoing sensemaking. We reframe text-to-3D interaction as negotiated meaning-making rather than linear prompting, and argue that effective systems require hybrid input modalities, transparent feedback, and low-cost iteration.

CCS CONCEPTS • •Human-centered computing~Human computer interaction (HCI)~HCI design and evaluation methods~User studies

## 1 INTRODUCTION

A user might describe an imagined space as "*a cozy mountain cabin at sunset, with a fireplace and bookshelves, overlooking a pine forest*." Recent text-to-3D generative AI systems can translate such descriptions into navigable virtual environments within minutes, without requiring traditional 3D modeling expertise. This capability marks a significant shift in how digital worlds can be authored, with implications for architectural visualization, game design, virtual collaboration, and immersive storytelling. However, unlike text-to-image generation—where alignment between intent and output can be assessed through immediate visual inspection—evaluating a generated 3D environment requires embodied exploration. Users must move through space, assess scale and spatial relationships, and experience whether the environment coherently supports the intended atmosphere. As text-to-3D systems transition from research prototypes to production platforms, a central HCI question emerges: how do users interact with prompt-based 3D world generation across the full lifecycle of ideation, creation, evaluation, and refinement?

Prior research on prompt-based generative AI shows that translating creative intent into prompts is cognitively demanding. Users engage in iterative trial-and-error, gradually developing mental models by observing outputs and reverse-engineering system behavior rather than specifying intent upfront [Liu 2022; Zamfirescu-Pereira 2023]. Human–AI interaction research further emphasizes the importance of expectation management, transparency, and meaningful human control in such systems [Amershi 2019; Shneiderman 2020]. Yet empirical HCI studies have largely focused on systems producing textual or two-dimensional outputs. Although technical surveys document rapid advances in text-to-3D generation methods [Lee 2024], empirical studies of how users actually interact with prompt-based 3D world generation remain scarce. Existing work primarily examines text-to-image models as ideation aids within traditional 3D pipelines rather than direct generation of explorable environments [Liu 2023]. As a result, we lack an interactional account of how users form, test, and revise spatial intent when evaluation itself is embodied.

Text-to-3D world generation introduces interaction demands that differ qualitatively from two-dimensional generative systems in three critical ways. First, evaluation is inherently embodied: users must navigate environments to judge spatial coherence, scale, and navigability rather than inspecting static outputs. Second, spatial structure and scale are central quality dimensions that are difficult to specify precisely through natural language alone. Third, iteration is costly: each refinement requires substantial processing time followed by renewed spatial exploration, amplifying the consequences of opaque feedback and unpredictable generation. Together, these characteristics suggest that interaction with text-to-3D systems cannot be understood as a linear prompt–output pipeline. Instead, users must continuously interpret system behavior, revise expectations, and adapt strategies over time.

Understanding text-to-3D interaction therefore requires a lifecycle-oriented perspective that foregrounds sensemaking. Users externalize evolving mental models through prompts, interpret probabilistic outputs under conditions of limited transparency, and evaluate results through embodied experience. Classic interaction theory highlights how users' mental models shape goal formulation, execution, and evaluation [Norman 2013], while prior work shows that intelligibility and explanation are critical for supporting understanding in intelligent systems [Kulesza 2013]. However, these frameworks largely assume interpretable feedback and carefully authored environments. In AI-generated 3D worlds, spatial coherence is probabilistic rather than designed, and evaluation unfolds through exploration rather than visual inspection. Immersive qualities such as spatial presence—extensively studied in designer-authored virtual environments [Slater 2009; Skarbez 2017]—may therefore manifest differently in generative contexts, where alignment between intent and output is uncertain and continually renegotiated.

To address this gap, we conducted a mixed-methods study examining user interaction with Marble, a commercial text-to-3D world-generation platform. The study was structured around the full lifecycle of AI-driven world creation. Using think-aloud protocols, we captured users' reasoning, expectations, and sensemaking across five interaction stages: idea formation, prompt formulation, system processing, world exploration, and prompt refinement [Ericsson 1993]. We complemented these qualitative data with validated quantitative measures of usability, presence, and engagement [Brooke 1996; Sánchez-Vives 2005], alongside open-ended reflections on prompt–output alignment and iteration strategies.

This work makes three contributions. First, we reframe text-to-3D world generation as a process of continuous, embodied sensemaking rather than linear prompting, showing how users develop intent through interaction with probabilistic spatial outputs. Second, we introduce two conceptual constructs that explain key interactional tensions in generative 3D systems: asymmetric expressibility, whereby users readily convey themes and atmosphere through language but struggle to specify spatial structure and scale; and episodic presence, in which brief moments of immersion emerge when expectations align with outputs but do not accumulate into sustained place illusion. Third, we demonstrate that iteration breakdowns arise primarily from structural interaction barriers—such as poor discoverability, opaque feedback, and high temporal costs—rather than from user limitations, reframing refinement failure as a system-level design problem. Together, these contributions establish text-to-3D interaction as negotiated meaning-making shaped by embodied evaluation, system transparency, and the costs of iteration.

## 2 LITERATURE REVIEW

### 2.1 Prompt-Based Generative AI: From Text-to-Image to Text-to-3D

Prompt-based generative AI has reshaped digital content creation by requiring users to translate creative intent into textual descriptions that guide system outputs. Research on text-to-image generation shows that this translation is cognitively demanding: users engage in iterative trial-and-error, struggle to predict how phrasing affects outputs, and often lack vocabulary for desired visual qualities [Liu 2022]. Extending this work, [Zamfirescu-Pereira 2023] demonstrate that non-AI experts approach prompt design opportunistically rather than systematically, struggling to develop effective strategies in ways that echo challenges observed in end-user programming and interactive machine learning.

Together, these studies establish prompt formulation as a process of interactional learning through which users construct mental models of how generative systems interpret intent, rather than a matter of language fluency alone.

Recent advances have extended prompt-based generation to three-dimensional content. Text-to-3D systems combine natural language processing, large language models, and generative 3D representations to translate prompts into spatial outputs. Prior work documents rapid technical progress, including improvements in visual fidelity, physical plausibility, and real-time performance [Wang 2024; Wang 2025], as well as broader surveys of neural representations and diffusion-based approaches [Lee 2024]. Computationally generated virtual environments have long supported applications in architecture, engineering, education, and cultural heritage [Chandramouli 2006; Forte 2024; Grant 2013]. Prompt-based text-to-3D systems, however, introduce a distinct interaction paradigm in which users author explorable spatial environments directly through language.

Despite these advances, understanding of human-centered interaction with prompt-driven 3D world generation remains limited. Existing work largely prioritizes algorithmic performance and visual realism, offering little insight into how users form mental models, refine prompts, or evaluate generated environments relative to intent. HCI research on AI-assisted 3D workflows has primarily examined text-to-image models as ideation aids within traditional modeling pipelines [Liu 2023], rather than direct text-to-3D generation in which users conceptualize, explore, and refine complete spatial environments through language alone. Moreover, prior studies typically assume outputs that can be assessed through immediate visual inspection. In contrast, text-to-3D systems render prompting an embodied evaluative process, where understanding emerges through navigation of three-dimensional space rather than static inspection. How interaction changes under these conditions—particularly when evaluation is embodied and iteration incurs substantial temporal and cognitive cost—remains underexplored, motivating our investigation of interaction across the text-to-3D lifecycle.

## 2.2 Mental Models and Interaction with Intelligent Systems

To understand how users engage with generative 3D systems, we draw on established interaction theory and recent work on AI-mediated creation. As shown in Figure 2, Norman's interaction cycle [Norman 2013] characterizes how users translate goals into actions and interpret system state to assess outcomes, identifying two critical gaps where breakdowns occur: the gulf of execution (difficulty mapping intentions to available actions) and the gulf of evaluation (difficulty interpreting system outputs relative to goals). This framework is particularly relevant for generative systems, where users must externalize ambiguous creative goals into representations the system can act upon, then evaluate whether outputs align with intent under conditions of uncertainty and limited transparency.

While Norman's interaction cycle provides a valuable lens for understanding interaction breakdowns (see Figure 2), it was developed for interfaces in which users select from known actions and receive relatively interpretable feedback. Prompt-based generative systems challenge these assumptions in fundamental ways: users specify intent rather than actions, system behavior is probabilistic rather than deterministic, and outputs may diverge across iterations. As a result, classic execution–evaluation framings under-describe interaction when outcomes are experiential, probabilistic, and spatial. Traditional notions of execution and evaluation therefore do not fully capture how users reason about system behavior or recover from mismatch in generative contexts.

Applying Norman's framework to prompt-based 3D generation requires reinterpreting what constitutes an "*action*" in interaction. Generative AI introduces intent-based outcome specification, in which users articulate what they want through natural language without specifying how the outcome should be produced [Weisz 2024]. The primary execution step is thus not direct manipulation but prompt formulation—encoding spatial intent, stylistic preferences, and implicit constraints into language. This challenge is compounded by generative variability, whereby outputs may differ in structure or detail even when input remains unchanged [Weisz 2024]. Users must therefore develop strategies for working with variability, reframing it as a property to be managed rather than eliminated. When results diverge from expectations, users enter iterative trial-and-error cycles to refine prompts, a dynamic previously observed in text-to-image systems [Liu 2022].

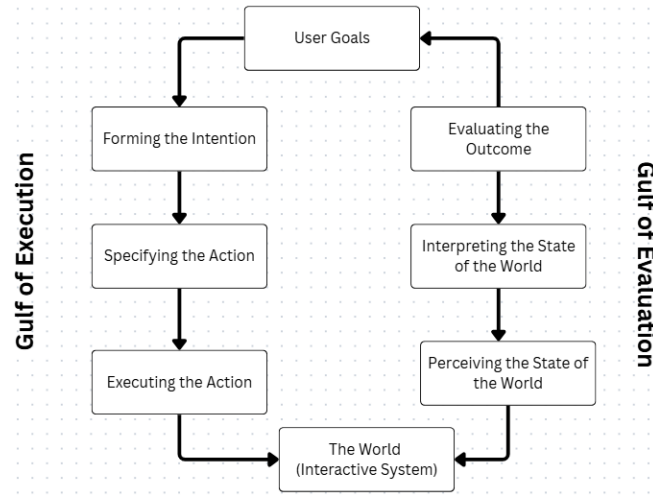

Figure 2: Norman's Interaction Cycle [Norman 2013]

From a broader HCI perspective, these dynamics align with traditions that conceptualize interaction as sensemaking and exploration rather than execution of predefined tasks. In such interactions, users iteratively interpret system behavior, revise goals, and construct understanding through cycles of action and feedback rather than through upfront planning. Prompt-based 3D generation foregrounds this mode of interaction, as users must learn how language maps to spatial outcomes by exploring, interpreting, and adapting to system responses over time.

Evaluation in prompt-based 3D generation also differs qualitatively from conventional interface feedback. Rather than verifying discrete outputs, users must perceive and interpret an immersive spatial environment—assessing layout, scale, navigability, and coherence—against mental models that may be only partially articulated. Prior work shows that users' ability to make sense of system behavior depends on transparency in how inputs map to outputs and on feedback mechanisms that support diagnosis when mismatches occur [Kulesza 2013]. In generative 3D systems, where prompt-to-output mappings are opaque and generation feedback is limited, these demands significantly amplify the gulf of evaluation.

Together, these perspectives motivate the need for a lifecycle-oriented account of interaction with prompt-based 3D world generation that extends Norman's interaction cycle [Norman 2013] to accommodate intent-based outcome specification [Weisz 2024], probabilistic generation, and experiential evaluation through spatial exploration. Synthesizing these perspectives, we propose an interaction model that foregrounds how mental models are externalized through prompts, revised through embodied exploration, and reshaped across iterative cycles under conditions of limited transparency and high iteration cost.

Figure 3 illustrates this interaction model for prompt-based 3D world generation. Rather than executing predefined actions, users externalize evolving mental models through prompt specification and refinement, expressing desired themes, atmosphere, and constraints in natural language. The generative system produces probabilistic 3D worlds, which users evaluate through embodied exploration of layout, scale, navigability, and coherence. This experiential evaluation feeds back into users' mental models through interpretation and sensemaking rather than simple success–failure assessment.

Iteration thus unfolds as a continuous loop of mental model revision, prompt reformulation, and spatial exploration, rather than as a linear prompt–output pipeline.

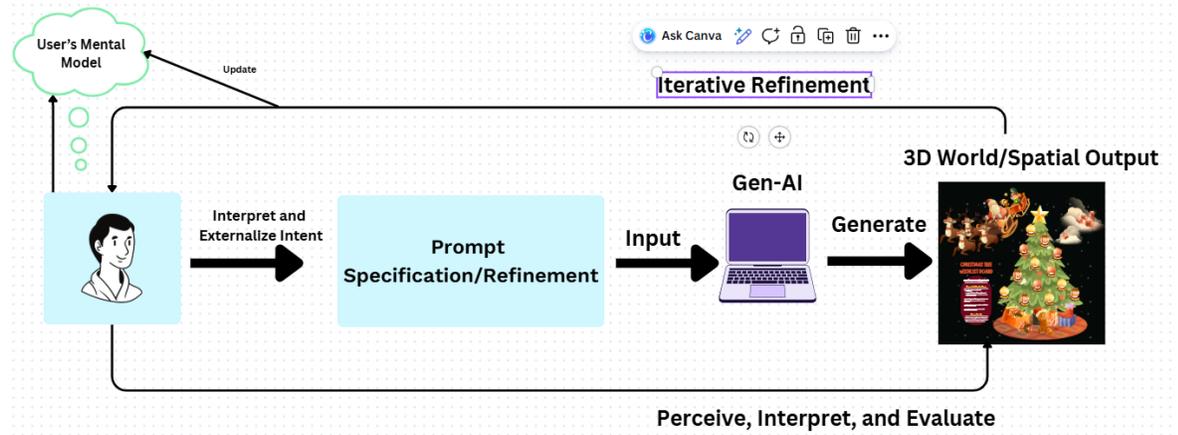

Figure 3: Conceptual Diagram of Prompt-to-3D Interaction Lifecycle

## 2.3 Usability

In generative AI systems, usability extends beyond interface mechanics to include how users formulate intent, understand system behavior, and iteratively refine outputs under uncertainty [Amershi 2019; Nielsen 1993; Norman 2013]. In prompt-based systems, where users express intent through natural language and evaluate variable outputs, understanding, control, and refinement are central to effective use. These challenges are especially pronounced in text-to-3D generation, where users articulate spatial intent and evaluate environments through embodied exploration rather than direct manipulation. As a result, usability directly governs whether iterative sensemaking can be sustained, since each interaction cycle entails both temporal cost and embodied evaluation. Recent systems such as WorldGen [Wang 2025] demonstrate technical feasibility—producing traversable 3D worlds with control over layout and scale—but prioritize architectural efficiency over empirical examination of how users formulate prompts, interpret outputs, or refine results.

Conceptual work further emphasizes that usability in generative AI shapes creative flow, perceived control, and users' ability to reason about system behavior. Misaligned language-model workflows can fragment design thinking and disrupt creative processes [Popescu 2023; Caramiaux 2025], while inaccessible interfaces prevent non-experts from translating creative intent effectively [Huang 2025]. Together, these perspectives underscore that usability in generative systems is integral to the generative process rather than an auxiliary concern.

Despite this recognition, prior work has paid limited attention to how usability challenges accumulate across iterative generation cycles. In text-to-3D systems, where each refinement may require substantial processing time and renewed spatial exploration, interaction friction does not merely slow interaction—it can terminate iteration altogether. Consequently, usability breakdowns shape whether users continue refining, abandon exploration, or reinterpret system capability. How such breakdowns emerge over time—and how they constrain sustained co-creative engagement—remains insufficiently understood, motivating our examination of usability as a dynamic, structural constraint on iteration across the interaction lifecycle.

## 2.4 Presence and Immersion in AI-Generated 3D Worlds

Presence refers to the psychological state in which users perceive themselves as located within a virtual space rather than the physical world, while immersion describes the extent to which a system perceptually surrounds users and supports three-dimensional interaction, thereby facilitating presence [Sánchez-Vives 2005; Skarbez 2017]. In traditional VR systems, presence is shaped by sensory fidelity, interactivity, spatial coherence, and narrative consistency—factors typically authored or designed in advance [Slater 2009].

These assumptions are challenged in AI-generated 3D worlds, where environments are produced through probabilistic rather than deterministic processes. Spatial layout, object relationships, and environmental coherence may vary across generations, introducing uncertainty that directly affects users' sense of presence. While presence has been extensively studied in designer-authored virtual environments, far less is known about how it emerges in prompt-driven, AI-generated worlds where users co-create environments through language and repeatedly evaluate evolving spatial outputs.

Existing theories largely conceptualize presence as a sustained experiential state. In contrast, generative 3D systems involve repeated transitions between creation, exploration, and refinement, during which spatial coherence evolves unpredictably across iterations. Under these conditions, presence may emerge intermittently rather than continuously, shaped by moment-to-moment alignment between user expectations and encountered spatial features. This reflects not diminished immersion, but a qualitatively different temporal structure of presence arising from iterative generative interaction. How such fluctuations unfold across interaction cycles—and how they intersect with usability and iteration costs—remains an open question motivating our empirical investigation.

## 2.5 Research Gaps and Questions

The preceding review reveals that while text-to-3D generation techniques have advanced rapidly, existing interaction frameworks offer an incomplete account of how users engage with these systems—particularly with respect to mental model formation, embodied evaluation, and iteration dynamics in probabilistically generated environments. In contrast to prior work that treats prompting as a specification or optimization task, understanding text-to-3D interaction requires examining how intent, interpretation, and experience evolve through use. To address these gaps, this study investigates the following research questions:

- RQ1: How do users conceptualize and articulate spatial intent when interacting with prompt-based AI systems for 3D world generation?
- RQ2: How do users interpret and evaluate AI-generated 3D environments relative to their expectations and mental models?
- RQ3: What interaction breakdowns and usability barriers arise during iterative prompt refinement and exploration?
- RQ4: How do experiential qualities such as presence and engagement fluctuate across interaction cycles in AI-generated 3D environments?

Collectively, these questions position text-to-3D world generation not as a prompt optimization problem, but as an interaction lifecycle shaped by continuous sensemaking and embodied evaluation.

## 3 METHOD

### 3.1 Study design, participants, and system

We conducted a mixed-methods user study examining interaction with prompt-based AI systems for 3D world generation, focusing on mental model formation, usability, iterative refinement, and experiential qualities such as presence and immersion. The study followed a single-condition exploratory design suited to under-theorized interaction dynamics, combining think-aloud protocols, structured observation, and post-task questionnaires to capture complementary qualitative and quantitative data. The procedure was structured around the interaction lifecycle described in Section 2.2 and followed a phased protocol comprising pre-test, in-test, and post-test stages to examine interaction over time.

Participants (N = 24) were recruited through university mailing lists and social media. Eligibility required participants to be aged 18–35, fluent in English, and familiar with general-purpose generative AI tools (e.g., ChatGPT), verified via a screening questionnaire. Participants were selected to ensure that observed breakdowns reflected challenges specific to text-to-3D interaction rather than initial exposure to generative AI. Prior experience with virtual reality, gaming, or 3D design tools was not required. Sample size was determined using data saturation principles appropriate for qualitative think-aloud studies [Braun 2021].

The study was conducted online using Marble [World Labs 2025], a commercial generative AI worldbuilding tool that produces navigable 3D environments from user inputs (see Figure 4). Although Marble supported multiple input modalities, interaction was constrained to text prompts to isolate prompt-based world generation and examine how users articulate and revise spatial intent through language. Participants interacted exclusively through the text-based prompt interface on desktop or laptop computers using Google Chrome; mobile devices were not permitted.

Generated worlds were explored using standard mouse-and-keyboard navigation controls, enabling assessment of spatial relationships, scale, and coherence. Marble supported iterative refinement through prompt modification and full regeneration but did not provide direct object manipulation, explicit layout constraints, or low-level control over geometry or generation parameters. System feedback during generation was limited, and generation times ranged from approximately 8 to 16 minutes per world depending on prompt complexity and system load. These conditions were consistent across participants and formed part of the interaction context under investigation. Sessions were supervised synchronously by a researcher who provided procedural guidance and technical clarification (e.g., navigation controls) without advising on prompt content or refinement strategies. Audio, screen activity, and interactions were recorded with participant consent. All participants provided informed consent, were compensated for their time, and the study was conducted in accordance with institutional ethical guidelines for human subjects research.

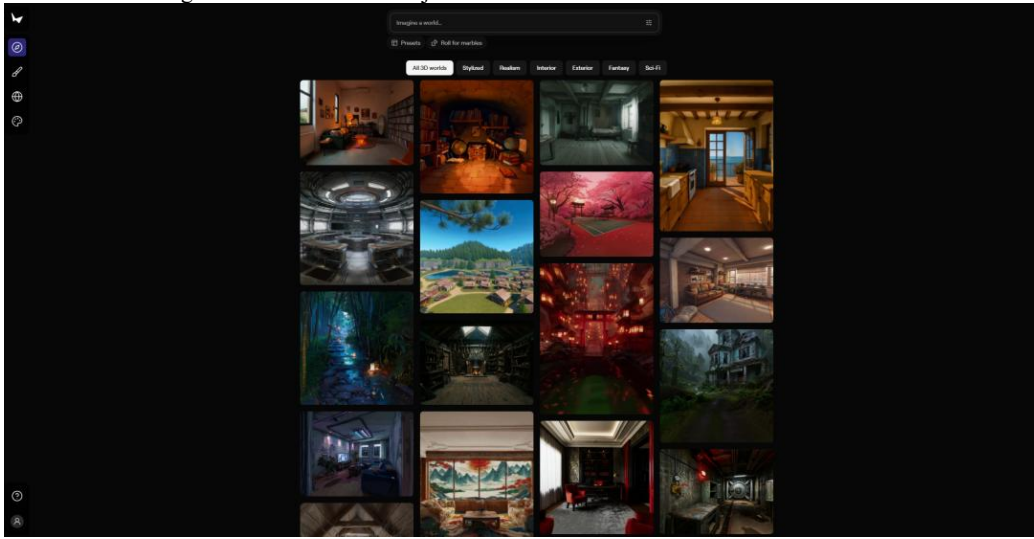

Figure 4: Marble System's Interface

## 3.2 Step by step design

The study followed a three-phase protocol comprising pre-test, in-test, and post-test stages, designed to capture interaction across the full lifecycle of prompt-based 3D world generation. This structure allowed us to examine how participants formed intent, interacted with the system, evaluated outputs, and decided whether to continue iterating.

During the pre-test phase (approximately 10 minutes), participants completed a demographic questionnaire assessing age, gender, educational background, and prior experience with generative AI, virtual reality, and 3D design tools. Participants then received a brief tutorial introducing Marble's core interactions: entering a text prompt, initiating generation, and navigating the resulting 3D environment. The tutorial focused exclusively on interface mechanics and navigation controls and did not include guidance on prompt content, wording, or strategy. This constraint was intended to minimize researcher influence on creative decision-making and to allow participants to independently develop prompting approaches during the task.

During the in-test phase (approximately 30–45 minutes), participants completed an open-ended worldbuilding task using a think-aloud protocol [Ericsson 1993] to capture reasoning, expectations, and interpretations throughout interaction. The task instruction was: "*Create a 3D world of your choice using text prompts. You may create any environment you wish and refine it until you are satisfied or choose to stop*." This framing supported creative freedom while maintaining a consistent task structure across sessions. Participants were free to determine the number of iterations they performed, allowing us to observe when and why refinement was sustained or abandoned.

The think-aloud protocol was organized around five interaction stages corresponding to the lifecycle framework introduced in Section 2.2:

- **Idea formation**: Participants described the world they intended to create and reflected on their confidence in translating that vision into natural language.
- **Prompt formulation and execution**: Participants verbalized how they decided which details to include in prompts and how they interpreted interface affordances.
- **System processing**: While awaiting output, participants reflected on their understanding of system status and expressed affective responses such as curiosity, uncertainty, or frustration.
- **World exploration**: Participants navigated the generated environment and articulated evaluative judgments concerning spatial layout, visual quality, coherence, and navigability.
- **Prompt refinement**: Participants who were dissatisfied revised and resubmitted prompts, explaining what motivated changes and what they expected revisions to improve.

Throughout the in-test phase, a researcher monitored interactions, recorded observational notes, and logged all prompt inputs and revisions. Researcher intervention was limited to procedural clarification (e.g., navigation controls or task reminders) and did not include suggestions regarding prompt content or refinement strategies.

During the post-test phase (approximately 10–15 minutes), participants completed questionnaires assessing usability, presence, engagement, and immersion. They also responded to open-ended questions addressing perceived prompt–output alignment, challenges encountered during refinement, and overall experience with the system. Figure 5 shows a participant interacting with the text-to-3D system during a study session.

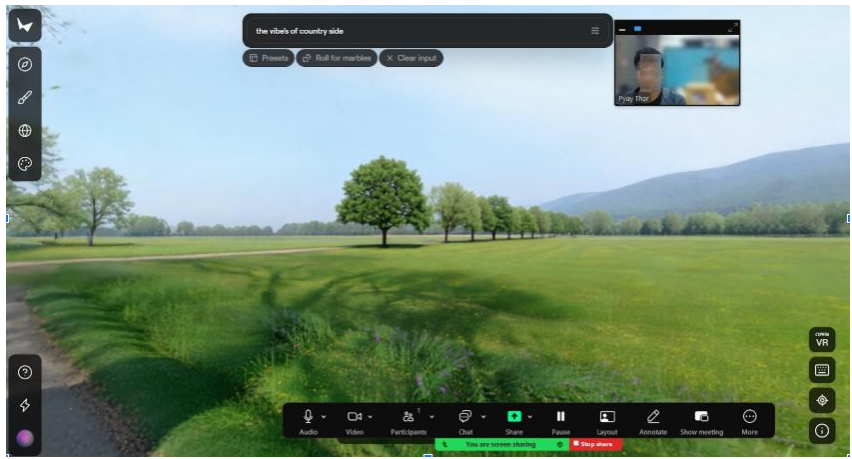

Figure 5: Screenshot of a Usability Experiment Session

### 3.3 Measures

Quantitative measures were collected via post-task questionnaires. Usability was measured using the System Usability Scale (SUS), a 10-item instrument assessing perceived effectiveness, efficiency, and satisfaction [Brooke 1996]. Presence was assessed using a six-item questionnaire adapted from place-illusion measures in virtual reality research [Sánchez-Vives 2005; Slater 2009], capturing participants' sense of "*being there*" on a 7-point Likert scale. Items addressed feeling present in the environment, perceiving the world as real, feeling "inside" the space, retrospective vividness, and whether the environment felt like a stable place; full items are provided in the Appendix. Engagement was measured using six items from the User Engagement Scale (UES), focusing on involvement, absorption, and focused attention [O'Brien 2010]. Immersion was captured using a single-item self-report measure on a 7-point scale, an approach common in immersive systems research [Skarbez 2017]. Internal consistency for multi-item scales is reported in the Results section. Qualitative data comprised think-aloud transcripts, screen recordings, researcher observation notes, and prompt logs collected during the in-test phase. Participants also responded to open-ended questions addressing overall experience, prompt–output alignment, challenges in expressing spatial intent, and factors influencing satisfaction and refinement decisions. These data provided insight into mental models, sensemaking processes, and experiential evaluations.

### 3.4 Data Analysis

All qualitative data—including think-aloud transcripts, screen recordings, observer notes, prompt logs, and open-ended responses—were analyzed using reflexive thematic analysis following Braun and Clarke's guidelines [Braun 2021]. This approach was selected for its suitability in examining interactional and experiential phenomena without predefined hypotheses, allowing themes to be generated through engagement with the data rather than imposed a priori. Analysis was primarily semantic, with higher-level interpretation developed during later stages of synthesis.

Analysis proceeded in four stages. First, two part-time researchers with relevant experiences conducted initial open coding on a subset of eight sessions to develop familiarity with the data and generate a broad range of candidate codes capturing participants' reasoning, observable behaviors, moments of uncertainty or breakdown, evaluative judgments, and affective responses. Coding was conducted independently to support analytic breadth, after which researchers engaged in iterative discussion to reflect on interpretations and develop a shared, evolving set of analytic codes. Consistent with reflexive thematic analysis, we did not calculate intercoder reliability, treating coding as an interpretive process refined through analytic dialogue rather than as a measurement procedure requiring statistical agreement [Braun 2021].

Second, we conducted pattern coding, grouping related codes into higher-level categories representing recurring interactional phenomena. These patterns were examined across participants to identify both commonalities and meaningful variation, with themes iteratively refined through team discussion.

Third, we performed temporal synthesis to examine how behaviors, interpretations, and experiential responses unfolded over the interaction timeline. The five interaction stages described in Section 3.2 served as a sensitizing framework rather than a deductive structure, with stage boundaries and transitions emerging empirically from the data.

Fourth, we conducted cross-stage abstraction to examine how challenges, strategies, and mental models carried over between stages, enabling analysis of how interactional dynamics accumulated across iterative cycles rather than occurring in isolation.

Throughout analysis, verbal reports were triangulated with screen recordings and prompt logs to ensure interpretations were grounded in observable interaction rather than self-report alone. Norman's interaction cycle [Norman 2013] informed interpretation as an analytic lens for reasoning about execution, evaluation, and breakdowns, but did not function as a coding schema or constrain theme development.

Quantitative data from post-task questionnaires were analyzed using descriptive statistics, including means, standard deviations, and internal consistency (Cronbach's α) where applicable. These measures were used to contextualize qualitative findings and characterize experiential patterns, rather than to test hypotheses or establish causal relationships.

## 4 RESULTS

### 4.1 Participants

The study included 24 participants (12 male, 12 female), predominantly young adults aged 18–34. Most participants were students, with smaller proportions identifying as working professionals or academic researchers. Participants reported limited prior experience with immersive technologies: the majority had no or beginner-level experience with virtual and augmented reality, with no participants reporting advanced expertise. In contrast, participants demonstrated substantial familiarity with general-purpose AI tools, with most reporting intermediate to advanced use and none reporting zero exposure. This profile—high AI literacy combined with limited immersive-technology experience—provides important context for later findings on embodied evaluation, navigation difficulty, and iteration breakdowns in text-to-3D world generation.

### 4.2 Asymmetric Expressibility

Across participants, we observed a consistent asymmetry in how intent could be expressed through language-based 3D world generation. Participants were generally able to articulate high-level semantic content—such as themes, moods, activities, cultural references, and object categories—with relative ease. In contrast, fine-grained spatial structure, including layout, scale, relational positioning, and precise control over composition, proved substantially more difficult to specify and predict through prompts. We refer to this pattern as asymmetric expressibility: *language afforded strong control over what a world was about, but limited control over how that world was spatially realized*.

*4.2.1 Expressible intent grounded in lived experience and culture.*

Participants' initial ideas were overwhelmingly grounded in personally meaningful contexts, including lived experience, seasonal events, cultural memory, media consumption, and professional identity. One participant noted, "*Today is *Christmas Day, I want to create something related to Christmas... a lobby with family laughing*" (P0001EMM), while another drew on media consumption: "*I watch K-drama about peaceful life... the story is about an actress living in the countryside*" (P0002EMM). Others proposed place-based concepts tied to cultural significance, such as "*the Ayeyarwady River and the Magway Myatthalon Pagoda because of their cultural, religious, and geographical importance*" (P0007EMM), or symbolic combinations of technology and spirituality—an "*AI robot praying Buddha*" (P0008EMM). Participants also drew on hobbies and professional domains: one "*enjoys attending concerts, so he wants to create a prompt related to designing a concert stage*" (P0001WY), while another with "*a background in civil engineering... attempted to create prompts related to building based on her professional knowledge*" (P0004WY).

These ideas translated readily into prompts that emphasized objects, atmosphere, and activities. Prompts frequently took the form of additive lists or short descriptive phrases (e.g., "*Christmas tree and lobby and family gathering*," "*sunset background, pine trees, waves crashing*," "*modern house on the tree like a bird's nest*"), which the system expanded into richly detailed environments. Across many cases, AI-generated outputs successfully captured the conceptual intent of these prompts, often elaborating beyond what participants explicitly specified by adding stylistic coherence, environmental detail, and narrative framing. This indicates that participants approached the system as a medium for expressing conceptual and experiential intent, rather than as a precision design tool requiring formalized ideation strategies.

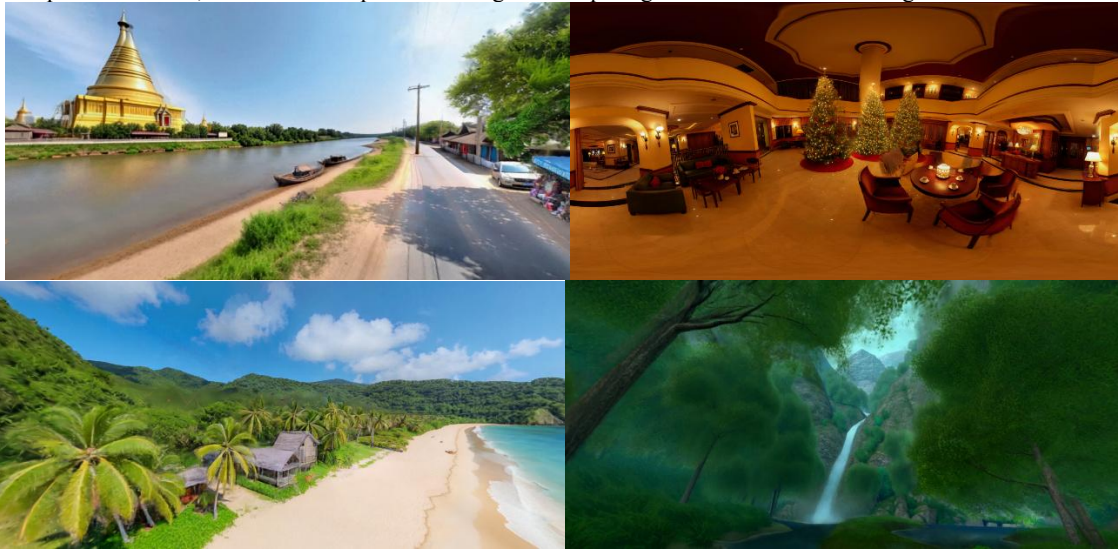

Figure 6: Selected Worlds Created by the Participants

*4.2.2 Underspecification of spatial relations and compositional control.*

While conceptual intent was readily expressible, participants consistently underspecified—or struggled to specify—spatial relationships and structural constraints. Raw prompt data show that even when participants refined prompts, changes primarily targeted semantic additions (e.g., more objects, people, cultural markers, time of day) rather than explicit spatial instructions. For instance, refinements often added elements such as "*Add Asian people and festival*" (P0002EMM), "*night time*" (P0001EMM), or "*add beach chairs and bungalows*" (P0001K), while leaving layout, scale, and relative positioning implicit.

When participants attempted to exert spatial control, results were mixed. One participant refined a prompt to specify "*only one table and some sofa. 3 peoples are sitting on sofa. I want night time*" (P0001EMM), yet the generated output did not reliably render the specified quantity or configuration. Another participant attempting manga-accurate character depictions found results "*completely different from prior expectations*" (P0003K), with the system favoring generic stylistic plausibility over domain-specific fidelity. Many participants reported that incremental refinements did not produce predictable spatial changes, prompting them to abandon fine-grained tuning in favor of rewriting prompts entirely. As a result, generated worlds often aligned well with participants' conceptual vision while diverging from their spatial mental models.

*4.2.3 Interface discoverability shapes early expression.*

Asymmetric expressibility was further shaped by early interaction demands. Idea formation did not occur as a clean, pre-interaction phase; instead, it unfolded alongside participants' efforts to understand where and how action was possible within the interface. Multiple participants experienced difficulty locating the prompt input box: one "*knows instantly to give a prompt but couldn't find the prompt input box*" (P0004K), while another "*took a while to find the prompt input box*" (P0002K). Observer notes indicated that "*the placement and visibility of the available tools were suboptimal, thereby reducing ease of navigation and discoverability*" (P0002WY). These discoverability challenges delayed expression not because participants lacked ideas, but because they were uncertain how to translate those ideas into system action.

Importantly, prior experience influenced confidence and fluency, not the nature of ideas themselves. One participant showed "*no confusion since she... has some background of graphics design*" (P0001EMM), while another was described as "*comfortable navigating the interface, experimenting with features, and translating her ideas into prompts*" (P0010WY). In contrast, novice users commonly reported uncertainty: one "*does not understand how to use them... feels confused by the interface and is unsure where to start*" (P0004WY). However, across experience levels, the content of ideas remained similarly grounded in personal meaning rather than technical abstraction, suggesting that asymmetry arose from the interaction model rather than differences in creative capacity.

*4.2.4 Linguistic confidence and interpretive uncertainty.*

Several participants expressed uncertainty about how language itself would be interpreted, particularly regarding sentence structure, grammar, and level of detail. One participant wanted "*to know write like story or also working with broken English*" (P0002EMM), while another had "a little bit struggle to create English sentence" and was uncertain "*how to construct sentences effectively*" (P0003EMM). A third participant questioned "*whether it was sufficient to generate a meaningful world*" (P0006EMM). Despite these concerns, raw prompt–output pairs show that the system often compensated for linguistic imprecision by generating coherent, detailed scenes, reinforcing participants' perception that semantic intent mattered more than linguistic form.

At the same time, this compensation contributed to asymmetry: because the system filled in spatial and compositional details autonomously, participants had limited visibility into how specific prompt elements mapped to spatial outcomes. This opacity made it difficult to learn how to reliably control structure through language alone.

*4.2.5 Curiosity sustains engagement despite asymmetry.*

Notably, asymmetric expressibility did not suppress engagement. Observational data consistently showed curiosity, anticipation, and motivation coexisting with confusion. Participants showed "*signs of anxious and curious*" (P0001WY), appeared "*highly curious and eager to explore the tools*" (P0002WY), and remained "*interested in learning... motivated to explore the features*" even when initially confused (P0011WY). One participant's "curiosity drives her to explore different features, experiment with new ideas, and observe how the system responds to each prompt" (P0010WY). This sustained engagement suggests that while asymmetric expressibility constrained precision, it did not undermine the system's appeal as a tool for exploratory worldbuilding.

Taken together, these findings characterize early interaction with prompt-based 3D world generation as a process in which users can readily express what kind of world they want, but struggle to specify how that world should be spatially organized. Asymmetric expressibility thus emerges as a structural property of language-driven 3D generation: it enables rapid, conceptually aligned world creation while limiting users' ability to precisely control spatial realization through prompts alone.

**4.3 Iteration Breakdowns are Structural**

Although many participants expressed clear intentions to refine and improve generated worlds, sustained iteration frequently broke down. These breakdowns were not primarily driven by lack of motivation, creative exhaustion, or insufficient ideas. Instead, they arose from structural characteristics of the interaction, including interface discoverability limitations, opaque system feedback during generation, long processing times, and resource constraints. Together, these factors disrupted the iterative loop between evaluation and refinement, even when participants had concrete plans for improvement.

*4.3.1 Discoverability barriers interrupt refinement momentum.*

A recurring cause of iteration breakdown was difficulty locating or reusing prior prompts and generated outputs. Multiple participants reported uncertainty about where refinement actions were available or how to retrieve previous prompts. One participant was "*confuse which button to get previous prompt*" and could not distinguish "*both previous prompt and AI generated prompt*" (P0001EMM). Others "*cannot find prompt refinement buttons*" (P0003EMM, P0004EMM) or experienced "*hesitation to regenerate since cannot find reuse box*" (P0007EMM, P0008EMM). In some cases, participants required external guidance—an observer noted, "*I tell her where the button is*" (P0004EMM). These discoverability barriers did not reflect indecision about what to change; rather, participants often articulated specific refinements—adding missing objects, clarifying quantities, adjusting style—but were unable to act on these intentions. Iteration stalled not at the level of ideation, but at the level of accessing the means of refinement.

*4.3.2 Opaque system feedback and prolonged processing undermine iterative control.*

System processing further contributed to iteration breakdowns by introducing uncertainty about system state and progress. One participant noted that "*due to the lack of real-time feedback or progress indicators, the user was unsure whether the system was still processing, had frozen, or had encountered an error*" (P0005WY). This uncertainty prompted compensatory behaviors such as "*repeated checking of other tabs and hesitation about whether to refresh or restart the process*" (P0004WY). Some participants "*missed the notification*" that generation had completed (P0002EMM, P0003EMM), while another "*was confused when he saw the 'world generated' notification but couldn't find the world*" (P0002K).

Processing times of 8–16 minutes, and in some cases exceeding 30 minutes (P0005EMM, P0006EMM), compounded this uncertainty. One participant "*felt that the processing time was excessively long... he became increasingly frustrated, as the delay disrupted his workflow*" (P0009WY). Another reported that "*the long generation time made her sleepy*" (P0005EMM). These temporal costs discouraged exploratory iteration, as participants became reluctant to test changes whose outcomes felt uncertain or delayed.

*4.3.3 Resource constraints truncate iterative exploration.*

Iteration was further constrained by visible resource costs. One participant "*was confused when he noticed that a single prompt used 2,000 credits and wanted to control or reduce this usage*" (P0002WY), while another could not continue because "*the third world could not be generated due to insufficient credits*" (P0002WY). These constraints introduced a cost–benefit calculus into refinement decisions: participants weighed the effort and waiting time of regeneration against uncertain improvement in output quality. As a result, iteration often ended not when participants were satisfied, but when structural limits were reached. One participant noted "*each generation taking a little bit too long*" while appreciating that "*5 worlds can be generated on a free account*" (P0001K), suggesting awareness of both temporal and resource boundaries.

*4.3.4 From incremental refinement to prompt rewriting.*

When incremental refinements failed to produce visible changes, participants shifted strategies. One participant's "*second world showed no visible changes compared to the first world*" (P0002K) despite targeted additions. When refinement appeared ineffective, participants abandoned fine-tuning: one "*wrote a new prompt for the third world by scratch*" and "*got exactly as he wanted*" (P0002K). Successful rewrites were associated with renewed confidence, while repeated mismatch led to skepticism—one participant "*was skeptical of the generated world... completely missed the main theme*" despite added details (P0003K). This pattern suggests that the system supported episodic successes rather than sustained, cumulative refinement. Iteration was viable when changes produced noticeable effects, but fragile when outcomes appeared opaque or inconsistent.

*4.3.5 Structural—not motivational—breakdowns.*

Across sessions, participants consistently demonstrated curiosity, engagement, and concrete ideas for improvement. Even when dissatisfied, they remained "*interested in learning... motivated to explore the features*" (P0011WY) and showed "*strong interest and anticipation... despite extended waiting periods*" (P0004WY, P0005WY). Iteration breakdowns therefore cannot be attributed to lack of motivation or creative intent. Taken together, these findings reframe iteration breakdowns as a design problem rather than a user problem. Sustaining iteration depends not only on expressive language capabilities but also on transparent system feedback, discoverable refinement pathways, and manageable temporal and resource costs. When these conditions are unmet, even motivated users with clear improvement goals are unable to sustain iterative engagement.

## 4.4 Interaction as Continuous Sensemaking

Across all stages of interaction, participants' engagement with the prompt-based 3D world generation system was characterized not by linear execution, but by continuous sensemaking. Rather than treating prompting, system processing, exploration, and refinement as discrete steps, participants repeatedly interpreted system behavior, revised their understanding of system capabilities, and adjusted expectations about how language translated into spatial outcomes. Sensemaking unfolded through action, observation, and reflection, shaping how participants learned to work with the system over time.

*4.4.1 Sensemaking during prompt formulation.*

Participants frequently treated early prompts as exploratory probes rather than final specifications. Several intentionally began with broad or minimal descriptions to observe how the system would respond before committing to more detailed inputs. One participant created a general prompt "*to become familiar with how the system responds to simple instructions*" (P0007WY), while another experimented with basic elements such as windows or lighting before attempting a full scene. This behavior suggests that prompt formulation functioned as a hypothesis-testing activity, where participants used generation as a way to infer the system's interpretive logic (see Figure 7).

Uncertainty about linguistic form further reinforced sensemaking. Participants questioned whether prompts should resemble narratives or fragmented descriptions, and whether grammatical accuracy mattered. One participant asked whether it was acceptable to "*write like story or also working with broken English*" (P0002EMM), while another struggled with sentence construction and was unsure how "*the system expects or how detailed the prompts should be*" (P0011WY). Rather than halting interaction, this uncertainty led participants to observe outputs closely and infer rules retrospectively based on system responses.

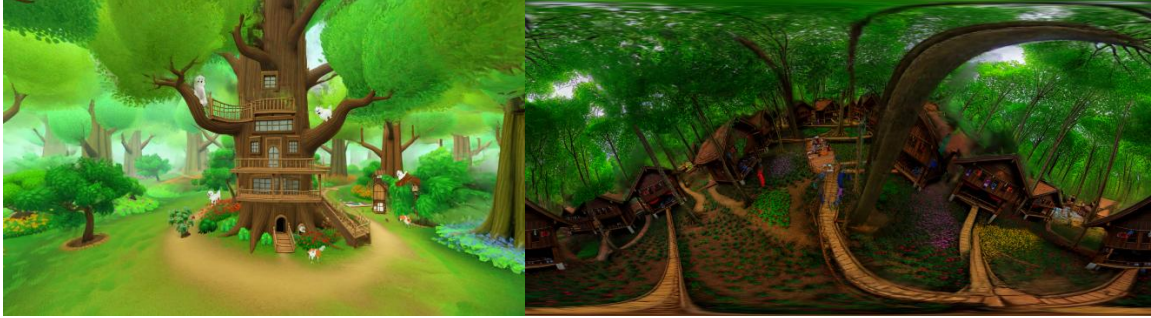

Figure 7: Original World (left) and Improved World (right)

*4.4.2 Sensemaking during system processing.*

System processing time functioned as an active interpretive phase rather than passive waiting. Although most participants recognized that generation was underway—"*Thinks the AI is drawing it right now*" (P0001K)—limited feedback and long delays prompted ongoing interpretation. Participants speculated about causes of delay, attributing it to internet stability, hardware limitations, or computational load (P0004EMM, P0009WY). Others compared the system to familiar generative tools, wondering whether it would "*generate AI-generated photos like other tools or mimic the original design*" (P0003K).

During this period, many participants actively explored the interface, examined other users' creations, or reflected on how prompts might be improved. One participant "*navigated around different tabs while waiting*" (P0001K), while another "*observed how to write prompt effectively*" by studying system behavior and outputs (P0002EMM). These behaviors indicate that sensemaking extended across temporal gaps, with participants using downtime to construct mental models of system operation.

*4.4.3 Sensemaking through embodied exploration.*

World exploration was a critical site of sensemaking, where participants evaluated prompt–output alignment through embodied navigation. Participants compared generated environments against their internal expectations, assessing realism, coherence, and spatial organization. When alignment was strong, participants inferred successful interpretation—"*it got the prompt pretty well, pretty much spot on*" (P0001K). When misalignment occurred, participants attempted to diagnose why, attributing discrepancies to missing detail, insufficient specificity, or system bias toward certain styles.

Navigation difficulties further shaped this process. Some participants were unsure how to move, pan, or zoom effectively, limiting their ability to fully inspect environments (P0002EMM, P0004K). These constraints affected not only experience quality but also participants' capacity to judge whether outputs truly matched intent. In response, participants often focused on salient visual features rather than full spatial layouts, reinforcing the tendency to reason about outputs at a conceptual rather than structural level.

*4.4.4 Sensemaking across refinement cycles.*

Prompt refinement represented an explicit attempt to operationalize sensemaking into control. Participants modified prompts based on perceived mismatches, adding constraints, quantities, or contextual details. However, when refinements failed to produce predictable changes, participants reinterpreted system behavior. Several concluded that fine-grained adjustments were ineffective and instead rewrote prompts entirely, treating each new attempt as a fresh experiment rather than a continuation of prior work.

Over time, participants' prompts increasingly reflected language patterns observed in AI-generated scripts, incorporating stylistic descriptors, spatial framing, and lighting terminology. This convergence suggests that participants learned how to prompt by adapting to the system's representational language, rather than the system adapting to theirs. One participant explicitly noted that longer and more specific prompts led to better results, indicating learned associations between prompt structure and output quality.

*4.4.5 Sustained engagement through interpretive adaptation.*

Despite uncertainty, sensemaking processes supported sustained engagement. Participants remained curious and motivated even when outcomes were unpredictable. Observers noted "*signs of anxious and curious*" behavior (P0001WY), "*eagerness to explore the tools*" (P0002WY), and continued experimentation despite confusion (P0011WY). Engagement was thus maintained not through mastery or efficiency, but through ongoing interpretation and discovery.

Taken together, these findings characterize interaction with prompt-based 3D world generation as an interpretive, sensemaking-driven process, in which users continuously negotiate meaning, control, and expectation. Rather than progressing linearly from idea to output, participants learned through cycles of action and reflection, constructing an evolving understanding of how language, interface affordances, and system interpretation jointly shape generated worlds.

**4.5 Episodic Presence in AI-Generated Worlds**

Participants' experiences of presence within AI-generated worlds were episodic rather than sustained, emerging in brief moments of spatial engagement rather than as a continuous sense of "*being there*." Both quantitative and qualitative data indicate that participants intermittently experienced spatial immersion when specific visual, semantic, or atmospheric elements aligned with expectations, but these moments were frequently disrupted by navigational difficulty, visual roughness, performance issues, or mismatches between intended and realized spatial structure.

*4.5.1 Quantitative evidence of episodic presence.*

Presence scores reflected moderate but unstable spatial immersion. On a 7-point scale, participants reported a mean presence score of $M = 3.92$ ($SD = 1.01$), indicating neither strong immersion nor complete detachment. Item-level analysis revealed a clear pattern: participants most strongly endorsed experiencing presence at certain moments ($M = 4.17$, $SD = 1.43$) and feeling a general sense of "*being there*" ($M = 3.96$, $SD = 1.43$), while reporting the weakest agreement with the statement that the environment felt like a stable place comparable to real-world locations ($M = 3.29$, $SD = 1.76$). This distribution suggests that presence emerged transiently, tied to localized impressions rather than sustained place illusion.

Correlation analysis further supports this interpretation. Usability was significantly correlated with presence (r = .465, p = .025), indicating that participants who experienced fewer interactional barriers reported stronger moments of spatial presence. However, presence was not significantly correlated with engagement, suggesting that episodic presence was not a prerequisite for continued interest or motivation. Given the modest sample size (n = 24), these correlational findings should be interpreted cautiously and warrant replication.

*4.5.2 Qualitative manifestations of episodic immersion.*

Qualitative observations and participant comments provide insight into how episodic presence was experienced. Moments of presence were often triggered by visual coherence, atmospheric detail, or conceptual alignment. Participants described being surprised or impressed when generated worlds closely matched their mental imagery—one participant "*became surprised and impressed by how accurately it reflects real-life situations*" (P0001WY), while another noted that the world "*got the prompt pretty well, pretty much spot on*" (P0001K). These moments were frequently accompanied by positive affect, such as delight, curiosity, or increased confidence in the system.

However, such moments were rarely sustained. Participants commonly reported that environments felt "*raw*," "*rough*," or incomplete upon extended exploration (P0003WY, P0005WY). Several noted missing elements—such as absent characters, insufficient detail, or unexpected stylization—that disrupted immersion. One participant remarked, "*I don't feel reality… because of the environment, which is more like in a movie*" (P0005), explicitly distinguishing engagement from a sense of presence. These disruptions indicate that presence depended on short-lived alignment rather than on enduring spatial coherence.

*4.5.3 Embodied interaction and breakdowns of presence.*

Navigation and performance constraints played a significant role in limiting sustained presence. Some participants struggled with basic embodied interaction, such as moving, panning, or zooming within the environment (P0002EMM, P0003K, P0004K). These difficulties constrained participants' ability to explore worlds fluidly, often shifting attention from the environment itself to the mechanics of interaction. As a result, presence was interrupted by moments of disorientation or effortful control.

Performance issues further contributed to episodic immersion. Participants reported lag during navigation and longer generation times for subsequent iterations, which reduced responsiveness and continuity. In several cases, participants expressed interest in deeper exploration but were discouraged by system latency or visual instability, limiting their capacity to remain immersed for extended periods.

*4.5.4 Presence as alignment-dependent rather than place-based presence.*

Taken together, these findings suggest that presence in AI-generated 3D worlds was alignment-dependent rather than place-based. Participants experienced brief moments of presence when visual output, atmosphere, and conceptual intent aligned closely, but these moments were fragile and easily disrupted by spatial inconsistency, interactional friction, or unmet expectations. Presence did not accumulate over time into a stable sense of place; instead, it emerged and receded across interaction stages.

Importantly, episodic presence did not determine overall engagement. Participants often remained motivated and curious even when presence was weak or intermittent, indicating that exploratory value and creative interest compensated for limited spatial immersion. This reinforces the interpretation that, in prompt-based 3D world generation, presence is not a continuous state but a situational experience, shaped by momentary alignment between user intent, system interpretation, and embodied interaction.

### 4.6 Usability and Engagement

Usability and engagement were assessed through standardized questionnaires and qualitative observations. Together, these data indicate a system that participants found moderately usable yet consistently engaging. Usability limitations shaped how easily participants could act within the system and influenced experiences of presence, while engagement was sustained by curiosity and exploratory value despite interactional friction.

*4.6.1 Usability outcomes and interactional variability.*

System usability was measured using the SUS. The mean SUS score was M = 64.02 (SD = 13.65), slightly below the commonly cited benchmark of 68 for average usability. Scores ranged widely (35–95), indicating substantial variability in participants' experiences. This dispersion reflects divergent encounters with interface discoverability, navigation controls, and system feedback during generation and exploration.

Qualitative data contextualize these scores. Participants frequently reported difficulty locating key interface elements, including the prompt input box, generation status indicators, and prompt reuse or refinement controls. One participant noted "*the tabs are confusing and I am not sure where to go first*" (P08), while others found that "*the prompt box was not easily visible or clearly labeled, causing confusion and hesitation*" (P0005EMM, P0006EMM). Navigation controls within generated worlds also posed challenges: one participant "*knows how to pan the camera but doesn't know how to move around*" (P0004K), while another reported "*the tool is a bit laggy to navigate*" (P0003K). These issues contributed to lower usability ratings among participants who experienced repeated breakdowns in basic interaction flow.

Despite these limitations, some participants—particularly those with prior experience in design or technical tools—reported smoother interactions. One showed "*no confusion since she... has some background of graphics design*" (P0001EMM), while another was described as "*comfortable navigating the interface, experimenting with features, and translating her ideas into prompts*" (P0010WY). This suggests that usability was not uniformly low, but unevenly distributed depending on participants' ability to infer system affordances and recover from breakdowns.

*4.6.2 Engagement as sustained interest rather than absorption*

Engagement was measured using six items from the UES. Participants reported moderate engagement overall (M = 3.47, SD = 0.63) on a 5-point scale. Item-level responses indicate that engagement was characterized primarily by sustained interest and motivation to continue, rather than deep absorption or flow. Participants most strongly endorsed statements such as "*I wanted to continue using the system*" (M = 3.83, SD = 1.17) and "*The experience was engaging overall*" (M = 3.75, SD = 0.85). In contrast, indicators associated with immersion and temporal dissociation—such as "*I lost track of time*" (M = 3.17, SD = 1.09) and "*I was absorbed in the interaction*" (M = 3.25, SD = 0.79)—were rated lower.

These quantitative patterns align with observational data. Participants frequently remained curious and motivated even when encountering confusion, delays, or mismatches between prompts and outputs. Observers noted that participants appeared "*highly curious and eager to explore the tools*" (P0002WY) and remained "*interested in learning... motivated to explore the features*" even when initially confused (P0011WY). One participant's "curiosity drives her to explore different features, experiment with new ideas, and observe how the system responds to each prompt" (P0010WY). This suggests that engagement was driven by exploratory and creative affordances rather than seamless interaction or immersive continuity.

*4.6.3 Relationships among usability, presence, and engagement*

Exploratory correlation analysis further clarifies the relationship between usability and engagement. Usability was significantly correlated with presence (r = .465, 95% CI [.07, .73], p = .025), indicating that participants who found the system easier to use also experienced stronger moments of spatial presence. In contrast, usability was not significantly correlated with engagement (r = .134, 95% CI [−.29, .51], p = .541), nor was presence significantly correlated with engagement (r = .262, 95% CI [−.16, .60], p = .215). Given the modest sample size (n = 24), these null correlations should be interpreted cautiously and warrant confirmation in larger samples.

This dissociation suggests that engagement was not contingent on either usability or sustained spatial immersion. Participants often remained engaged even when usability was marginal or presence was episodic. Several participants explicitly distinguished between feeling immersed and finding the experience valuable: one noted, "*I don't feel reality… because of the environment, which is more like in a movie*" (P0005EMM), yet still described the system as engaging. Others characterized it as "a strong creative accelerator" (P15) that "*helped to generate ideas quickly*" (P18), emphasizing ideational value over experiential polish.

*4.6.4 Engagement persists despite usability constraints.*

These findings show that while usability constraints shaped interaction smoothness and experiences of presence, they did not uniformly diminish engagement. Instead, engagement was sustained by curiosity, novelty, and the ability to rapidly generate and explore conceptually rich worlds through language, even in the presence of interactional friction, performance issues, or interface confusion. Together, these results indicate that usability and engagement are partially decoupled in prompt-based 3D world generation: usability limitations constrained efficiency, predictability, and immersion, while engagement remained resilient, supported by the system's capacity to transform abstract ideas into explorable environments.

## 5 DISCUSSION

This study examined user interaction with prompt-based AI systems for 3D world generation across ideation, execution, evaluation, and refinement. We show that text-to-3D interaction is a continuous, sensemaking-driven process shaped by spatial embodiment, interface structure, and system transparency, in which users progressively construct and revise intent through iterative engagement with system outputs. Our findings yield three core insights: asymmetric expressibility, whereby high-level semantic qualities are easier to articulate than fine-grained spatial structure; structural iteration breakdowns, driven by interaction costs and system opacity rather than limited creativity or motivation; and episodic presence, in which immersion in AI-generated 3D worlds is momentary rather than sustained. Together, these insights position text-to-3D systems not as prompt-optimization tools, but as interactional systems that require support for sensemaking, embodied evaluation, and low-cost iteration.

*5.1.1 Conceptualizing and Articulating Spatial Intent (RQ1).*

Our results show that users conceptualize spatial intent mainly at a semantic and experiential level, focusing on themes, moods, activities, and culturally meaningful elements while leaving spatial relations, scale, and layout implicit. This extends prior findings from text-to-image generation, where users struggle to anticipate how language maps to visual outcomes and instead learn effective prompting through observation rather than deliberate planning [Liu 2022]. Text-to-3D generation, however, adds a distinct challenge: users must describe not only what should appear, but also how space should be structured and navigated.

We describe this tension as asymmetric expressibility—a structural property of language-driven 3D generation in which high-level semantic intent is easy to articulate and often well realized, while fine-grained spatial control is difficult to specify or predict using natural language alone. Participants typically refined prompts by adding semantic elements such as objects, atmosphere, or time of day, even when dissatisfied with layout or scale. This pattern reflects not a lack of prompting skill, but a mismatch between how people naturally describe imagined spaces and how generative systems translate language into spatial form.

This construct generalizes beyond the specific system studied. Marble's interaction model—a language-primary text prompt without direct manipulation, explicit spatial constraints, or layout templates—reflects interaction patterns common in many current commercial text-to-3D systems (e.g., Luma AI, Blockade Labs). The observed asymmetry stems from fundamental properties of natural language, which excels at conveying meaning, narrative, and affect but offers limited means for encoding precise spatial relations, proportions, and constraints. While hybrid modalities may mitigate this issue, language-primary interfaces under current design paradigms are likely to encounter this tension.

As a result, current text-to-3D systems are better suited for conceptual worldbuilding and rapid ideation than for precision-oriented spatial design. This positions text-to-3D generation as a divergent creative tool rather than a replacement for traditional 3D modeling, with important implications for system design and positioning.

*5.1.2 Interpreting and Evaluating AI-Generated Environments (RQ2).*

Evaluation of AI-generated 3D environments was inseparable from embodied interaction. Unlike text-to-image generation, where alignment can be judged through immediate visual inspection, 3D evaluation required movement, navigation, and spatial inhabitation. Participants assessed realism, coherence, and intent alignment by traversing environments, with expectations formed during prompting often revised as they encountered scale, spatial relationships, and missing or unexpected elements.

This process aligns with and extends [Norman 2013]'s interaction cycle. In generative 3D contexts, both the gulf of execution and the gulf of evaluation are amplified: users must express spatial intent through underspecified language and then interpret outputs through embodied exploration rather than direct visual comparison. Evaluation therefore becomes iterative and provisional, shaped by movement through space, distinguishing text-to-3D from other generative modalities and rendering static quality metrics (e.g., FID scores, rendering fidelity) inadequate proxies for perceived alignment.

Participants' experiences of interpretation drift, especially in stylistically constrained domains such as anime-inspired or character-specific worlds, further illustrate this challenge. Outputs could be visually plausible yet violate domain-specific expectations, prioritizing generic coherence over fidelity to established conventions. This echoes findings on intelligibility in intelligent systems [Kulesza 2013]: meaningful interpretation depends not only on transparency but on respecting constraints central to users' mental models. In spatial generative systems, plausibility alone is insufficient; users expect adherence to the structural and stylistic commitments implied by their prompts.

*5.1.3 Iteration Breakdowns as Structural Costs (RQ3).*

A central finding is that iteration breakdowns were structural rather than cognitive. Participants were motivated and had clear ideas for refinement, but iteration often failed due to interactional barriers such as poor discoverability of controls, opaque system feedback, long generation times, and visible resource costs. This challenges the assumption that ineffective iteration mainly reflects limited prompting skill [Zamfirescu-Pereira 2023]. Users typically knew what they wanted to change but could not act on it, as each iteration required substantial time (8–16 minutes) and full re-exploration, far exceeding the low-cost iteration of text-to-image systems.

Usability friction also directly undermined presence and experiential quality, not just user satisfaction. This aligns with human–AI interaction guidelines emphasizing visible system state, efficient iteration, and low exploration cost [Amershi 2019]. In text-to-3D systems, where iteration is both time-intensive and embodied, delayed or opaque feedback prevents users from forming causal models of how prompts affect spatial outcomes, reducing both control and learning. The implication is that designers should focus less on teaching users to "*prompt better*" and more on reducing iteration friction through architectural support such as partial regeneration, version comparison, and transparent prompt-to-output mappings.

*5.1.4 Episodic Presence and the Decoupling of Engagement (RQ4).*

Participants reported moderate presence and engagement, but results revealed a distinct pattern we term episodic presence: immersion arose briefly when visuals, atmosphere, and expectations aligned, and dissipated when navigation issues, performance problems, or spatial inconsistencies intervened. Item-level analysis supported this pattern, showing momentary feelings of spatial presence alongside lower ratings of inhabiting a stable, place-like environment.

This extends [Slater 2009]'s concept of breaks in presence. In authored VR, breaks disrupt an otherwise continuous sense of place [Slater 2009]. In contrast, AI-generated worlds showed the inverse: presence was intermittent, emerging only when conditions aligned rather than persisting by default. This suggests generative environments require different presence-oriented design strategies.

Engagement, however, persisted even when presence was weak or intermittent. We found no clear relationship between presence and engagement, or between usability and engagement, indicating that continued use was driven by curiosity, creative interest, and the novelty of seeing imagined worlds realized. This dissociation implies that generative 3D systems can sustain meaningful engagement without strong presence, and that improving one will not necessarily improve the other; distinct design interventions are likely needed.

The four constructs identified in this study are not independent but form a reinforcing cycle. Asymmetric expressibility ensures that spatial intent remains underspecified, leading to iteration attempts that structural barriers render costly and ineffective. The resulting worlds exhibit episodic presence—alignment emerges momentarily but does not stabilize. Users respond through continuous sensemaking, developing workarounds rather than solutions. This cycle suggests that addressing any single construct in isolation will be insufficient; effective text-to-3D systems must intervene at multiple points—expanding expressibility through hybrid input, reducing iteration costs, and supporting sensemaking through transparent feedback.

*5.1.5 Contributions.*

These findings yield four contributions to HCI research on generative AI and spatial interaction. First, we provide a detailed empirical characterization of user interaction with text-to-3D systems across the full lifecycle from ideation through refinement, establishing baseline experiential benchmarks and documenting recurring interaction breakdowns. Second, we introduce asymmetric expressibility as a conceptual construct capturing the fundamental tension in language-driven spatial generation: users readily convey high-level themes and atmosphere but struggle to specify fine-grained spatial structure through natural language alone—an asymmetry rooted in the language-to-space translation problem rather than user skill deficits. Third, we characterize episodic presence as a distinctive experiential quality of AI-generated 3D environments, where immersion emerges momentarily when expectations and outputs align rather than persisting as sustained place illusion, suggesting that presence in generative spatial contexts requires design approaches distinct from those developed for authored VR environments. Fourth, we reframe iteration failure as structural rather than motivational, providing empirical evidence that users abandon refinement because interface barriers interrupt the iteration loop—not because they lack creative intent—thereby shifting design responsibility from training users toward building supportive interaction ecosystems. Together, these contributions highlight a broader implication: realizing the potential of text-to-3D generation requires moving beyond prompt-centric interfaces toward interaction ecosystems that support ongoing sensemaking, exploration, and refinement.

*5.1.6 Design Implications.*

Together, these findings highlight clear priorities for future text-to-3D system design. While some emerging systems—including Marble—are beginning to support additional input modalities, our results specifically inform language-primary, prompt-driven interaction and hybrid workflows in which text remains the dominant means of specifying intent. The implications below address how such systems can better support sensemaking, embodied evaluation, and iteration.

First, systems should support hybrid interaction modalities to address asymmetric expressibility, as natural language alone is insufficient for precise spatial specification. Text input should be complemented by mechanisms that externalize spatial intent without requiring full manual modeling, such as spatial templates, visual constraint overlays for layout and scale, and post-generation in-world adjustments that preserve semantic intent.

Second, system processing should be treated as an active interaction phase rather than passive waiting. Given generation times of 8–16 minutes, systems should support ongoing interpretation through progressive previews, explicit progress indicators with estimated completion times, interruptibility, and parallel generation that allows exploration of prior outputs during processing.

Third, iteration must be architecturally low-cost. Instead of requiring full regeneration for each change, systems should support partial regeneration, version branching for comparison, and prompt differencing that visualizes how input changes affect spatial outcomes. These mechanisms can support causal learning and sustain iterative engagement under uncertainty.

Finally, navigation should be designed as evaluation infrastructure rather than a neutral utility. Because evaluation in text-to-3D systems depends on embodied exploration, systems should provide multiple navigation modes with clear affordances, guided viewpoints, and bookmarking or annotation tools to support systematic comparison across iterations.

*5.1.7 Implications for Generative AI Research Beyond 3D.*

Although this study focused on text-to-3D world generation, several findings extend to complex generative AI systems more broadly. Sensemaking-driven interaction is likely to generalize across modalities in which outputs are complex and cannot be assessed at a glance: users develop intent while discovering system affordances, adapt strategies through observation, and iteratively revise evaluations. This pattern may apply to domains such as text-to-video, generative music, and multi-step AI agents, suggesting that the proposed interaction lifecycle is useful beyond 3D.

More broadly, structural iteration barriers scale with output complexity. As generative outputs become richer and generation times increase, iteration breakdowns intensify, creating a design imperative to evolve interaction architectures that sustain viable iteration loops. Finally, asymmetric expressibility may characterize other language-mediated generative systems, where semantic intent is easy to convey but fine-grained structural specification remains difficult. Examining this tension across modalities could support a more general theory of language-mediated generative interaction.

*5.1.8Limitations.*

This study has several limitations that constrain the generalizability of its findings. First, it examined interaction with a single text-to-3D system (Marble). Although Marble reflects many affordances of current commercial platforms, interaction patterns may differ across systems, particularly those that emphasize alternative workflows or tighter integration of hybrid input modalities. In addition, while Marble supported multimodal inputs at the time of the study, interaction was intentionally constrained to text-based prompting in order to isolate language-driven world generation. Future comparative work should examine how hybrid inputs—such as image references, sketches, or 3D assets—reshape expressibility, iteration, and sensemaking dynamics.

Second, the study captured only short-term interaction, leaving longer-term learning trajectories unexplored. With extended use, users may develop more effective prompting strategies, more accurate expectations, or different relationships to iteration cost. Longitudinal studies are needed to examine how mental models evolve over time and whether challenges such as asymmetric expressibility diminish with expertise.

Third, participants were primarily young adults from academic settings with high general AI literacy but limited immersive technology experience. This profile may limit applicability to expert 3D designers, VR practitioners, or users from different linguistic and cultural backgrounds. Future work should examine how professional expertise, domain conventions, and cultural context shape interaction with generative 3D systems.

Fourth, presence and engagement were measured retrospectively rather than continuously, potentially masking moment-to-moment fluctuations during navigation and exploration. Future studies could employ experience sampling, behavioral proxies, or physiological measures to better capture the temporal dynamics of immersion in AI-generated environments.

Fifth, the lab-based, think-aloud protocol may have influenced prompting behavior by encouraging explicit reflection and verbalization. While appropriate for studying sensemaking, ecological studies of naturalistic, unsupervised use would complement these findings. Finally, the study focused on single-user worldbuilding; collaborative generation introduces coordination, authorship, and negotiation challenges that warrant separate investigation.

Despite these limitations, the study provides both an empirical foundation and a conceptual framework for understanding interaction with spatial generative AI, underscoring the importance of continuous sensemaking, low-cost iteration, and hybrid expression as text-to-3D systems enter creative, collaborative, and immersive workflows.

# 6 CONCLUSION

This study shows that prompt-based 3D world generation is a negotiated, co-creative process shaped by continuous sensemaking, spatial embodiment, and interaction design. Rather than reflecting deficits in user creativity, breakdowns arise when systems fail to support the iterative translation of intent into spatial experience. The findings introduce two concepts that distinguish spatial generative AI from two-dimensional modalities: asymmetric expressibility, in which high-level themes and atmosphere are readily conveyed through language while fine-grained spatial control remains difficult; and episodic presence, where brief moments of immersion emerge without accumulating into a sustained sense of place. We further show that iteration breakdowns stem from structural interaction barriers—such as poor discoverability, opaque refinement pathways, and high temporal costs—rather than from limits in user motivation. Effective generative 3D systems therefore require not only capable models, but interaction ecosystems that support transparent intent articulation, embodied exploration, and low-cost iteration. As text-to-3D technologies enter creative workflows and immersive applications, attending to the human experience of worldbuilding will be essential to realizing their full potential.